\documentclass[referee]{aa}
\usepackage{graphics}
\usepackage{psfig}

\begin{document}


\title{Proper Motions and CCD-photometry of Stars in the Region of the Open Cluster Trumpler\,2}
\author{V.N. Frolov\inst{1} \and J.K. Ananjevskaja \inst{1}
\and E.G. Jilinski \inst{1,2} \and D.L. Gorshanov \inst{1}}

   \offprints{V.N.Frolov, The Main Astronomical Observatory,
Pulkovo, St.\,Petersburg, Russia. email: vfrol@gao.spb.ru}

   \institute{The Main Astronomical Observatory, Pulkovo, St.\,Petersburg,
Russia
 \and  Observat\'orio Nacional / MCT, Rio de Janeiro, RJ, Brazil}

\date{ }

\authorrunning{Frolov}

\titlerunning{Open Cluster Trumpler\,2}

\abstract

\maketitle


\begin{abstract}  -The results of the complex study of  galactic open
cluster Trumpler\,2 are presented.The positions of approximately
3000 stars   up to the limit magnitude $B\sim16.25$ mag in the
area $80\arcmin\times80\arcmin$ around the cluster were measured
in order to obtain the  proper motions. Measurements of star
positions on 6 plates with the maximal epoch difference of 63
years were taken by means of the automated measuring complex
"Fantasy". The root-mean error of the relative proper motions is
$4.2\,mas\,yr^{-1}$.  The catalogue of BV magnitudes of all the
stars in the investigated area was compiled. Astrometric selection
of the cluster members within  the region of $R<16\arcmin$ from
the center of the cluster was made by means of the W.Sanders
method(1971).  In that field 192 stars were found to have the
individual membership probability greater then 85$\%$, 148 of them
are situated within the $\pm3\sigma_{B-V}$ band around the main
sequence (MS) of the cluster. They are considered to be cluster
members by two criteria. The U-B$\sim$B-V diagram plotted  for the
astrometrical cluster members  by the data taken from the
Washington catalogue of the UBV photometry in the galactic cluster
fields (Hoag et al. 1961) has made it possible to redefine the
value of the color excess E(B-V)$=0.^m$40 instead of previously
published values ranged from 0$.^m$30 to 0$.^m$34.  The
superposition of the MS of the cluster with the  ZAMS
Schmidt-Kaler (1965) for thus accepted E(B-V) leads to the
coincidence at the value of the apparent distance module
$(V-M_V)=10.50$ which corresponds to the distance $r= 725 pc$. The
luminosity and mass functions of the Trumpler\,2  were constructed
and the value of the slope ($\Gamma=-$1.90$\pm$0.22) was
determined. The cluster age of 8.913$\times10^7$yr was determined
on the base of Girardi et al. (2000) isochrone grid. Correspondent
to this age the turn-off point of the MS is (B-V)$_0=-0.^m$16. It
is shown that the red giant on the late stage of the evolution (st
N.1095) belongs to cluster and indicated the brightness
variability. The possibility that the number of both known and
recently discovered variables are cluster members was considered.
The catalogues of the proper motions and of the photometry of
stars are appended. \footnote{Tables~ 2,3,3A,5 will be only
available in electronic form at the CDS}
\end{abstract}

\section{Introduction}
 The open cluster Trumpler\,2 is located near the famous rich clusters
h and  $\chi$  in the Perseus constellation. Its equatorial and
galactic coordinates are: $\alpha=2^{\rm h}37^{\rm m}15^{\rm s}$,
$\delta=+55\degr59\arcmin$; $\ell=137\fdg 4$, $b=-3\fdg89$
$(2000.0)$. In the Tr\"umpler (1930) catalogue it is classified as
II 2p (a pure cluster  with little central concentration). All
that was known about the cluster to the present day is based on
the data from the  Washington catalogue "Photometry of stars in
Galactic cluster fields"(Hoag et al. 1961)(HJ). For Trumpler\,2 it
included UBV photoelectrical photometry of 28 stars in the
$15\arcmin\times15\arcmin$ area and photographic photometry of 133
stars in the circle  $R<22\arcmin$ around the supposed cluster
center. All the contemporary determination of the cluster
parameter are based on that data (Fig.\,1 a\&b presented the
Color-Magnitude diagram(CMD) and the two-color diagram of the
cluster from that study). The existence of the cluster itself was
not confirmed by any other original papers. This area was never
investigated kinematically on the base of proper motions. Our
paper has to fill this gap.


\begin{figure} [t]
\centering{
\vbox{\psfig{figure=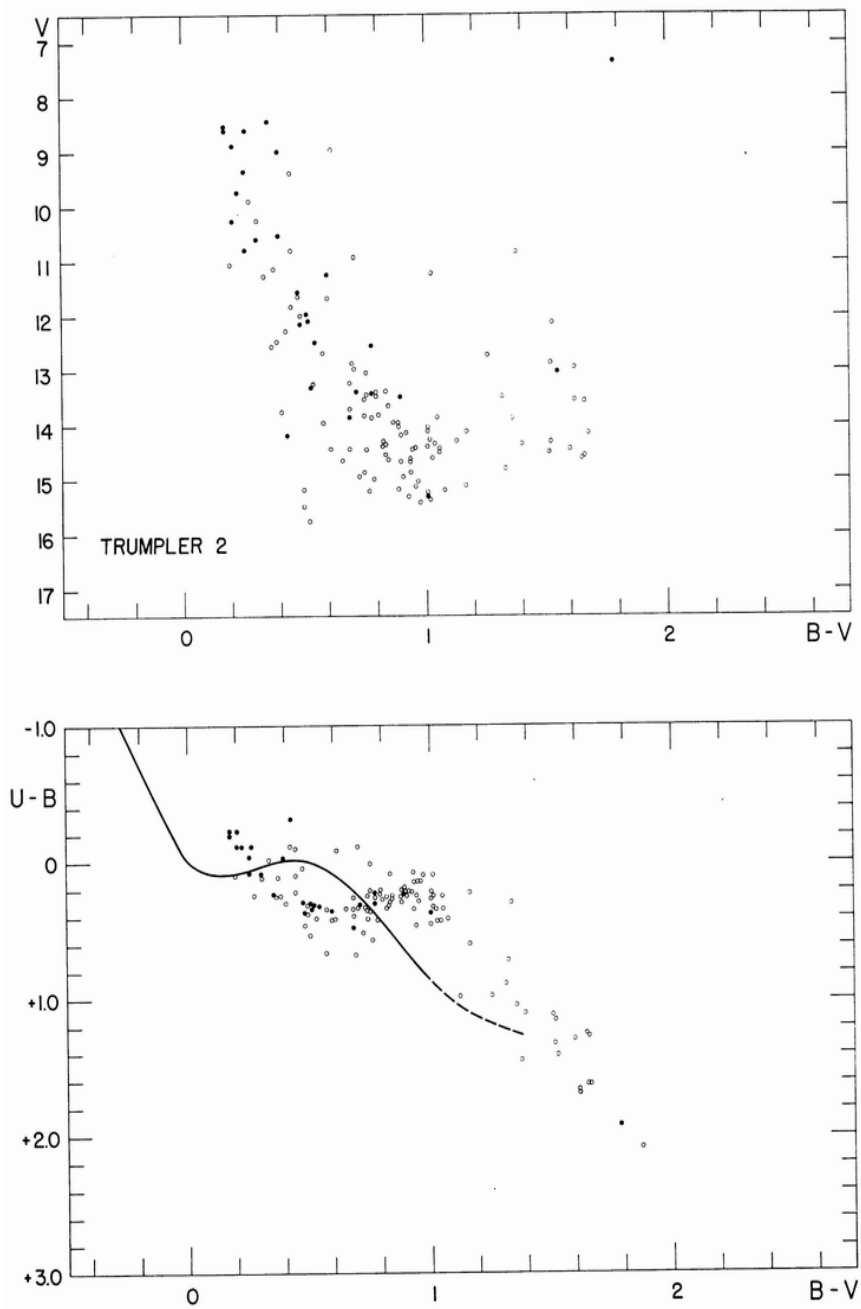,width=12cm,height=18cm}}\par }
\caption []{Color-magnitude and two-color diagrams of
  Trumpler\,2 from the Washington Catalogue. }
\end{figure}

\section{Astrometry}
\subsection{Proper motions}
The observational material belongs to the Normal astrograph
collection of the Pulkovo observatory and dates from 1937 to 2000
(Table\,1). The scale of the Normal astrograph plates  is ${\rm
60\arcsec\,mm^{-1}}$. The cluster region star positions from the
USNO-A 2.0 catalogue were used as an additional plate (epoch
1954.8).

\begin{table}
      \caption[]{Astrometric plates}
\begin{tabular}{lcll}
\hline
 Plate   &   Exposure  & \,\,\,Epoch  &      Quality\\
          &    (min)     &         &        \\
\hline \multicolumn{4}{c}{Early  epoch}\\
 K 376  &    25 & 1937 Sept. 12 & high \\
 D 80 & 30 & 1949 Mar. 31 & good \\
 D 135 & 30 & 1949 May 5 & poor \\
              \multicolumn{4}{c}{Recent  epoch}\\
 18160 &   25 &   1999 Dec. 12  &      poor \\
 18171 & 30 & 2000 Feb. 4 & good \\
 18175 & 25 & 2000 Feb. 6 & good \\
\hline
\end{tabular}
\end{table}

All of the plates were scanned by means of the automated measuring
complex "Fantasy" of the Pulkovo observatory. The description of
the complex was presented in Frolov et al (2002)(FJA). The
positions of the 2362 stars  in the  area
$80\arcmin\times80\arcmin$ centered on the cluster Trumpler\,2 up
to the limit magnitude $B=16.25$ mag were measured. Positions of
some faint stars were not measured on all plates because of  the
poor quality of their images. So from three to seven stellar
positions were used to derive proper motions(PM). PM were
determined by the line-method. The details of this method were
presented in Jilinski et al.(2000). The deepest plate $K376$ with
the faintest images was considered to be the central plate. It was
orientated by means of the stars positions from  the "Tycho-2"
catalogue. Then all the other plates were reduced to this central
one.  For that reduction were selected 72 reference stars in the
interval between $13.5$ and $14.5$ mag. The rms errors of derived
PM are: $\sigma _x = \pm 4.1\;{\rm mas\,yr^{-1}}$, $\sigma _y =
  \pm 4.2 \;{\rm mas\,yr^{-1}}$. The analysis of error distributions
demonstrates that they increased from $3.9$ to $4.2 \;{\rm
mas\,yr^{-1}}$ with the increasing of the distance from the area
center and from $3.1\;{\rm mas\,yr^{-1}}$ for stars with the
magnitude $B~<~10.5$~ mag to $4.4$ for $B>16$ mag stars. The
astrometrical results are given in the Table\,2 which columns
contain: 1 - the star number, 2 - the star number from the USNO-A
2.0 catalogue, 3,4 - its rectangular coordinates X and Y on the
plate K376, 5 - radius-vector, 6,7,8,9 - X and Y relative proper
motions and their errors, 10 - the number of used plates, 11 -
photographical B-magnitude, resulted from the "Fantasy"
measurements, 12 - notes(bl marks blends, bad - bad measurement,
abs -absence of the star on the plates, BD and AD - star
identification).

\subsection{Member segregation.}
   The selection of cluster members was made for stars located in
the circle with the $r<16\arcmin$ around the cluster center. That
region contains 313 stars with determined proper motions. Blended
images and stars measured only on two or less plates were
expelled. The size of this circle was selected according to two
reasons: first, it should be large enough for not to miss the
cluster members from the corona, second, not be too large to
litter the vector point diagram (VPD) with a large amount of the
field stars. We used  the method in which membership
probabilities, magnitude equations (ME) and parameters of the star
distributions on the vector point diagram  (VPD) were determined
step by step in iterative process. The details of  this procedure
are described in FJA. The starting selection was realized in the
circle with the R$<10\arcmin$ centered on the Trumpler\,2.  As a
first approximation stars with the photoelectric BV magnitudes
from the HJ  were considered to be the cluster members. Based on
proper motions and magnitudes of these stars the magnitude
equation (ME) was studied and the proper motions of all stars were
corrected for it. Then at each step characterized by increasing
value of R (+2$\arcmin$) the parameters of the bivariate Gaussian
distributions of proper motions, circular for the cluster and
elliptical for the field, were computed by the Sanders method
(1971). The preliminary selection of the cluster members had been
made in each iteration step. Then the proper motions were tested
for the ME. If it had been  revealed then all the PM in the
catalogue were refreshed. The latest step was made for the stars
in the circle with the radius R$<16\arcmin$. It is important to
note that it is impossible to divide  stars of  the cluster and of
the field  on the VPD visually because of the proximity of the
centers of their distributions. This fact is frequently registered
for the far clusters (for example see J. Sanner et
al.(199,2000,2001). Nevertheless it takes only 8 iteration to
reach the stable solution numerically with relative precision
E=0.001. The following initial values of the parameters were
accepted: $N_c/N=0.5$, centers of the distributions: the field -1,
1, the cluster: 1,-1, the dispersion: the field 12, 8, the cluster
4. Calculations resulted in the following values: the relative
number of cluster members and all stars $N_c /N = 0.74$; centers
of distributions: the  field $\mu_{x_f} = 0.98$, $\mu_{y_f}=
-2.11$ ; the cluster $\mu_{x_c}= 0.33$ , $\mu_{y_c}= -0.53$ ;
standard deviations: the field $\Sigma_{x}=  14.62$, $\Sigma_{y}=
9.28$, the cluster $\sigma_c=  4.78$.  All the values are in ${\rm
mas\,yr^{-1}}$. They were used to calculate the individual
membership probabilities of the stars in the area by the formulae:
 $$ P\left(\mu_x,\mu_y\right) =
\frac{N_c\Phi_c \left(\mu_x,\mu_y\right)} {N_c\Phi_c
\left(\mu_x,\mu_y\right) + N_f\Phi_f
\left(\mu_x,\mu_y\right)}\,\,\,\,,$$
 where $N_c$ - normalized number of cluster stars,
     $N_f$ - normalized number of field stars,
$\mu_{x_i}$, $\mu_{y_i}$ - proper motion in $x$ and $y$ for the
$\mbox{\it i}^{th}$ star.

 The resulting histogram shows that the stars with the probabilities
$P>85\%$ can be considered the cluster members (Fig.\,2). The
whole amount of such stars is 192.  The final VPD is presented in
Fig.\,3a. Cluster members are plotted by the solid dots, field
stars - by the crosses.
\begin{figure} [t]
\centering{
\vbox{\psfig{figure=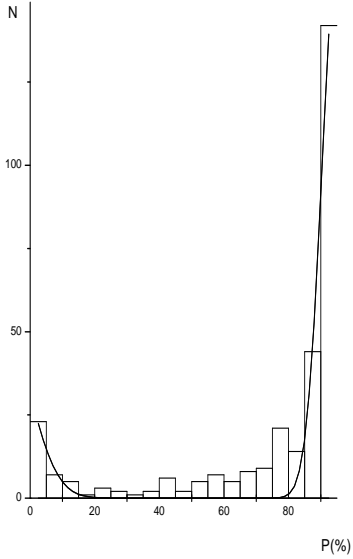,width=6cm,height=9cm}}\par }
\caption []{Final histogram of star membership probabilities.}
\end{figure}
\begin{figure} [t]
\centering{
\vbox{\psfig{figure=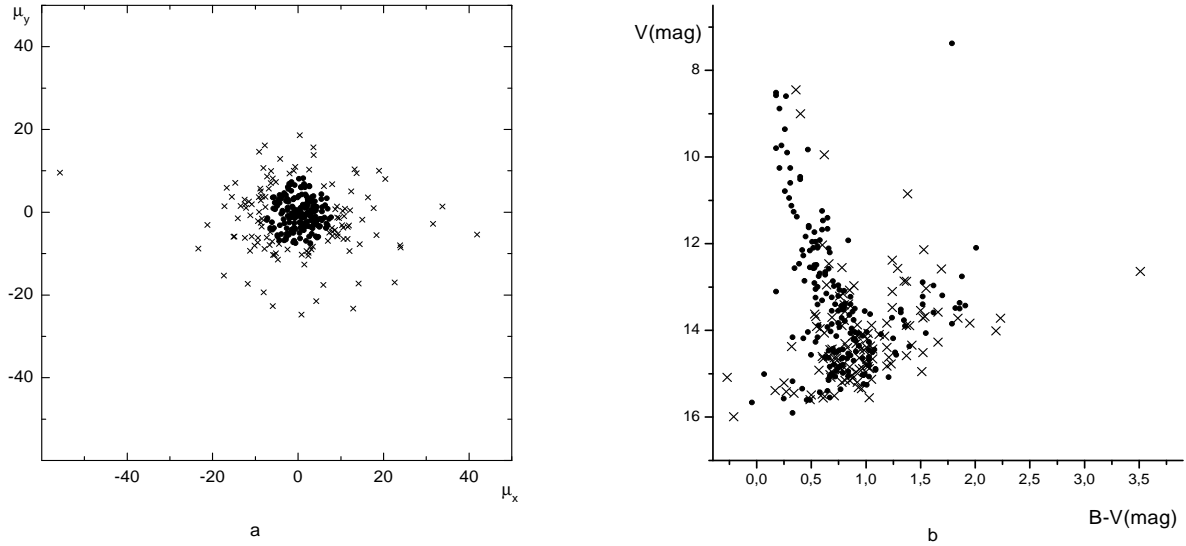,width=17cm,height=9cm}}\par}
\caption []{a) VPD and b) CMD of the central part (R$<16\arcmin$)
of Trumpler\,2. Astrometrical cluster members are plotted by solid
dots, field stars - by crosses.}
\end{figure}
\begin{figure} [t]
\centering{
\vbox{\psfig{figure=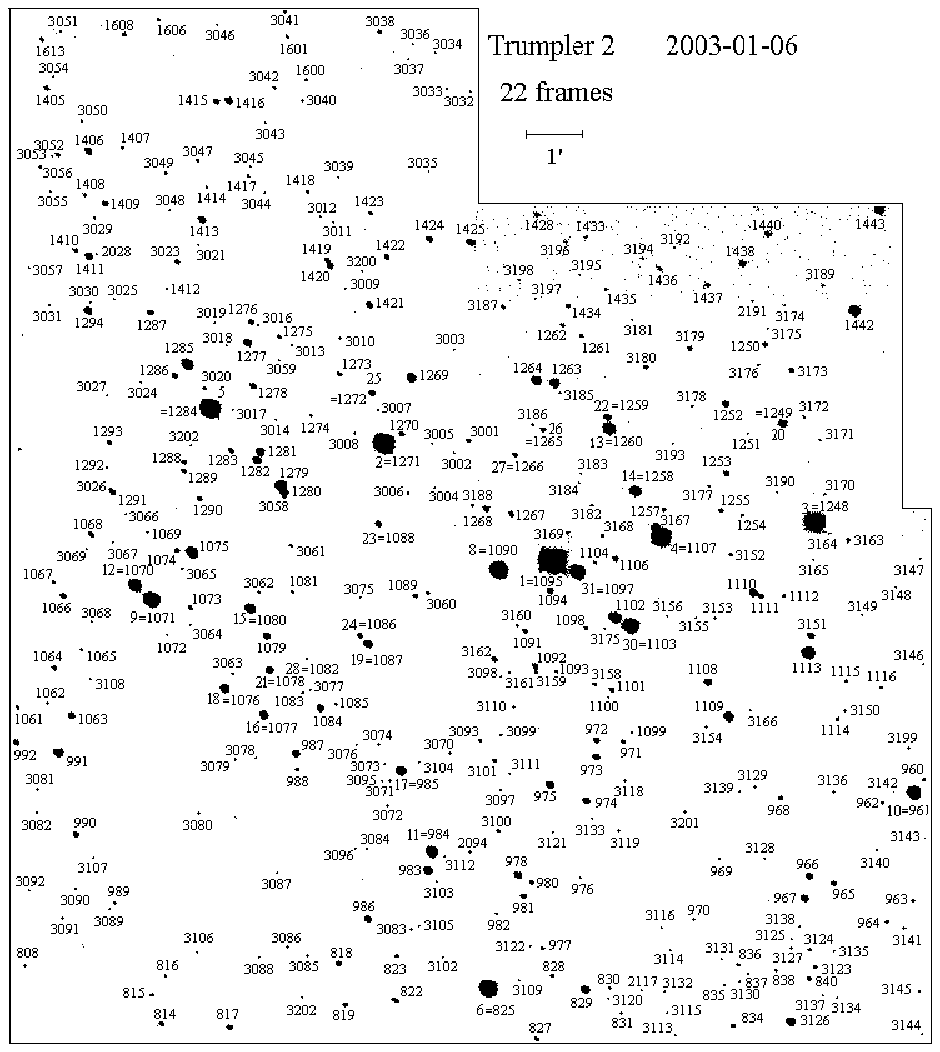,width=13cm,height=15cm}}\par}
\caption []{CCD map of the central part of the cluster
Trumpler\,2. The numbers of stars are given according to the
photometric study.}
\end{figure}

\section{Photometry}

The final conclusion of the star membership was based on two
criteria: its proper motion and its position on the CMD. Only
small part of situated in the central circle stars has got the
initial UBV-photometry(HJ). The $B$ and $V$ magnitudes of the rest
stars were determined partly as by-product of astrometric plate
measurements (image diameters) or by our original CCD-observations
and partly were calculated from the infrared J,H,K magnitudes from
2MASS catalogue by the derived transformation formulae.
Unfortunately due to the technical reasons we were able to obtain
CCD-magnitudes only in the area of $17\arcmin\times24\arcmin$. The
identification map of the central part of the cluster is presented
in Fig.\,4. The star number from the Washington catalogue is
indicated in front of the author's number. Data available from the
all different sources of the broadband photometry for each star
are presented in Table\,3, which columns contain: 1 - the star
number according to Table\,2, 2 - the radius-vector,3 - the
specific numbers from 1 to 10 corresponding to  photometry data,
3,4- B and V  magnitude,5 - B-V, 6 - notes. The description  and
correspondent comments of the specific numbers which mark the
sources of photometry are given in in Table\,4, which columns
contain 1 - the specific number (S/N), 2 - photometry band, 3 -
source of the material, 4 - the area dimensions in arcmin, 5 - the
photometric errors in magnitudes. \setcounter{table}{3}
\begin{table} [!]
\caption[]{Cross-references for the photometry}
\begin{tabular}{cccll}
\hline
 S/N  &   Bands  & Sourses  & Area &  Errors \\
 1  &    $B, V$  &      HJ, photoelectric magnitudes &  $17\arcmin\times17\arcmin$ &   $\sigma_V  = \pm0.021$ mag,
 $\sigma_{B-V} =\pm0.018$ mag\\
 2  &    $B, V$  &      HJ, photographic magnitudes  & $30\arcmin\times12\arcmin$  &  $\sigma_V = \pm0.014$ mag,
  $\sigma_{B-V}= \pm0.021$ mag\\
 3  &    $B_T, V_T$ & Tycho catalogue  &  &  $\sigma_V= \pm 0.060$ mag,\\
 &&& & $\sigma_{B-V}= \pm 0.100$ mag (all  stars),\\
   &             &  & & $\sigma_V= \pm 0.012$ mag,\\
   &&&& $\sigma_{B-V} = \pm 0.019$ mag ($V_T<9$ mag)\\
 4  & $B$ & Author's  photographic  magnitudes  &  $120\arcmin\times120\arcmin$ &  $\sigma_B= \pm0.09$ mag  for
 "red" stars\\
   &  & catalogue  &    &  $\sigma_B= \pm 0.03$mag  for "blue" stars\\
     &    $V$ &  Author's       CCD magnitudes   & $19\arcmin\times24\arcmin$ & $\sigma_V= \pm 0.06$ mag \\
     &          &   catalogue     & &\\
5  & $B$ & Author's photographic  magnitudes  &
$120\arcmin\times120\arcmin$ & $\sigma_B= \pm0.09$ mag  for
 "red" stars\\
   &  & catalogue   &    &  $\sigma_B= \pm 0.03$mag  for "blue" stars\\
 &  $V$     &   2MASS catalogue &  &    $\sigma_V= \pm 0.11$ mag \\
     &          & calculated  from the  J,H,K magnitudes & &\\
 6  & $B$ & Author's  photographic  magnitudes  &  $120\arcmin\times120\arcmin$ &  $\sigma_B= \pm0.09$ mag  for
 "red" stars\\
   &  & catalogue  &    &  $\sigma_B= \pm 0.03$mag  for "blue" stars\\
     &   $V$ & No data & &\\
 7   &   $B$ & USNO 2A catalogue & &\\
     &   $V$ & No data & &\\
 8   &   $B$ & Author's CCD magnitudes  & $19\arcmin\times24\arcmin$ & $\sigma_B=0.08$ mag \\
 && catalogue && \\
     &   $V$ & Author's  CCD magnitudes & $19\arcmin\times24\arcmin$ & $\sigma_V=0.05$ mag \\
     &     &           catalogue & &\\
 9   & $B$  & calculated from the values of $B-V$ & &\\
     &    & derived from the 2MASS  catalogue & &\\
     & $V$  & Author's    CCD magnitudes  & $19\arcmin\times24\arcmin$ & $\sigma_V=\pm 0.06$ mag \\
     &    &                 catalogue & &\\
 10  & B  & derived from the  2MASS  catalogue & &\\
    & V   &   Author's    CCD magnitudes &   $19\arcmin\times24\arcmin$  &
    $\sigma_V =\pm 0.05$ mag \\
          & &               catalogue & &\\
\hline
\end{tabular}
\end{table}

Notes for specific numbers:\\
 1 and 2.  While  plotting the characteristic curve $V$
the systematic difference between photoelectric and photographic
values $(V_{pe}-V_{pg})=0.^m01$ in HJ catalogue was detected. It
was taken into account. The same difference was detected for the
colors B-V.\\
 3. The real precision of the photometric data given in
Tycho - 2 catalogue for stars $V_T> 9$ mag is much lower then
claimed. That's why  we preferred alternative sources for these
stars.\\
 4, 5, 6. $B$ - photographic values of the major quantity of stars
in the studied region were determined by measuring their diameters
by means of the "Fantazy" complex. It is well known that $B$
magnitudes obtained with the Pulkovo Normal astrograph are
infected by the color equation(CE). That's why characteristic
curves $B\sim F(D)$ were plotted separately for "red" $(B-V\geq
1.0)$ and "blue" $(B-V < 1.0)$ stars. $B$ magnitudes of  the
"blue" stars were not corrected for CE as it's value was
non-significant (smaller than the mean error). At the same time
from original observations made of the "red" stars the value of
the correction was 0.35 mag per magnitude and was taken into
account in all cases.\\
 4. $V$- magnitudes were obtained from original observations made with
the 320 mm mirror astrograph ZA-320 of the Pulkovo Observatory
equipped with the CCD-receiver ST-6 with a fixed TC241 matrix
(Bekyashev et al.(1998)).\\
 5. $V$ -magnitudes were calculated with the J,H,K
magnitudes from the infrared 2MASS  catalogue by the derived
transformation formula :
 $$ V=0.02528+1.04259J-0.83847(J-H)+3.16435(J-K)+2.39154(J-H)^2
-1.34984(J-K)^2 $$
 7. $B$-magnitudes from the USNO-2A  catalogue were treated only as control
   values.\\
 8. $B,V$ magnitudes were obtained from original observations  with the 320 mm mirror
astrograph ZA-320 of the Pulkovo Observatory equipped with the
CCD-receiver ST-6 with a fixed TC241 matrix(Bekyashev et
al.(1998)).\\
 9,10. $B$ -magnitudes were calculated with the J,H,K magnitudes from
the infrared 2MASS  catalogue by the derived transformation
formulae: $$
B=1.0621J+1.0732(J-H)+2.9973(J-K)+0.3785(J-H)^2+0.4259(J-K)^2$$
 $$ B-V=-0.21236+0.04164J-2.51276(J-H)+2.43958(J-K)+2.27583(J-H)^2
-0.14797(J-K)^2 $$

Table\,3A contains the additional magnitudes $B$ and $V$ obtained
by means of CCD-photometry for stars which brightnesses are less
than the limit sensibility of the astrometric  plates. Their
numbers begin with the 3 on the identification map  of the central
part of the cluster (Fig.\,4). Columns of the Table\,3A contain:
 1 - the star number, 2,3,4 - B-magnitude, its error and the
number of observations, 5,6,7 - V-magnitude, its error and the
number of observations. As the standards for these observations we
took Washington photoelectric and partly (for the faint stars)
photographic photometry. Errors of the magnitude determination
were defined as external relative to the standards. Photometric
data from the Table\,3 were used to construct the CMD for 313
stars in the circle R$<16\arcmin$.  As it was mentioned in the
process of membership determination the positions of stars on CMD
were used as a second parameter. Because of unequal accuracy of
the material the following application priority order of the
specific numbers (S/N) was set: 1(photoelectric HJ) - 28 stars,
2(photographic HJ) - 56 stars, 3 (Ticho) - 6 stars, 8 (CCD) - 95
stars, 4(the combination of authors photographic and CCD)- 1 star,
5 - (the combination of authors photographic and 2MASS) - 127
stars. CMD of all these stars is shown in Fig.\,3b.

\section{Analysis and conclusion}

 Stars of the IV and V luminosity classes falling on the CMD into
the $\pm 3\sigma_{B-V}$  interval from the cluster main sequence
were  considered to be the photometric cluster members. As the
average error of the color index determination rises from the
bright to faint stars approximately from 0.02 mag to 0.11 mag the
width of the MS band increases.

In order to determine the mean absorption in the region of the
cluster the two-color diagram $(U-B)\sim(B-V)$ was plotted using
the HJ $UBV$ photometry only for the 44 astrometric cluster
members with $P\geq 85\%$ falling on the MS. For the inclination
$E(U-B)/E(B-V)=0.72$ the best coincidence of the line of normal
colors  with the two-color sequence of the cluster was achieved at
the value of the color excess $E(B-V)=0.40$ mag (Fig.\,5) which
differs from  accepted in existing compiled catalogues values:
0.30(Hoag,1966) and 0.34 (Electronic Database WEBDA,
2001\footnote{http://obswww.unige.ch/webda}). With thus fixed
value of color excess the superposition of the MS of the cluster
with the ZAMS (Schmidt-Kaler,1965) was made. During this procedure
the special attention was paid to the stars with photoelectric
photometry. The best coincidence was achived at the value of the
apparent distance module $(V-M_V)=10.50$ which corresponds to the
true module $(V-M_V)_0= 9.30$  and the distance $r= 725 pc$. In
Fig.\,6  CMD diagram for probable cluster members (according to
both photometric and astrometric criteria) is presented. Solid
circles marks the stars supplied by photoelectric photometric data
and hollow circles - all the other.

The catalogue of the stars in  the central circle
(R$<16\arcmin$)is presented in Table\,5. Its columns contain 1 -
the author's number, 2 - sign of membership, 3,4 -the rectangular
coordinates of star(arcmin), 5  - the distance from the center
(arcmin), 6,7,8,9 - the proper motions and their root-mean-square
errors($mas\,yr^{-1}$), 10 - the number of the astrometric plates,
11 - the membership probability, 12 - the photometry specific
author's number (see Table\,4), 13,14 - the star magnitude and
color.
\begin{figure} [t]
\centering{
\vbox{\psfig{figure=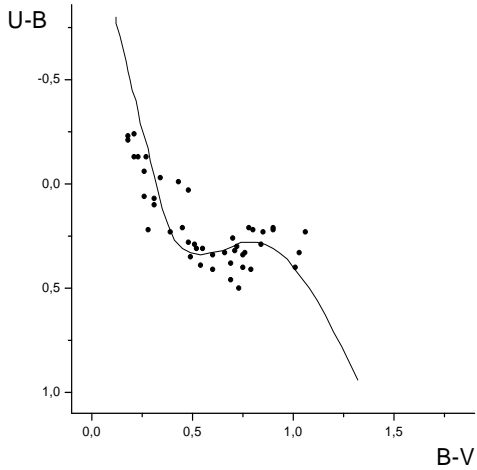,width=7cm,height=7cm}}\par}
\caption []{Two-color  diagram of the cluster Trumpler\,2.}
\end{figure}

Located in the cluster's center  red giant (st. N 1095)  has high
individual membership probability. Large dispersion in its
photometric data may be the evidence of the variability. Another
red giant (st. N 1748), though it does not fall in the 16\arcmin
circle where the individual membership probabilities were
determined, relates to the third class of Eibighausen (1942)
classification. It means that  this star lies within a circle of
the $R=2\sqrt2\sigma$ on the VPD where 98\% of all cluster members
are situated (Kadla,1966). The same refers to the st.~N 1941 which
can have travelled far from the cluster center but is
gravitationally linked with it. These possible cluster members are
marked by "prob". In  Fig.\,6 they are plotted  by $\triangle$.

\begin{tabular}{lccccl}
 $N$  &Radius-vector  &  $  M_V$  &   $(B-V)_0$ &    $ P$ &  Notes\\
      & (arcmin)& mag & mag & \% &\\
 1095  &1.15&   -3.12   &    1.39  &        94 &  suspected variable star\\
 1748  &17.35 &  -2.94   &    1.48   &       prob & \\
 1941  &36.70 &  -3.19    &   0.51    &   prob & \\
\end{tabular}

 The estimation of the age of the cluster is based on the grid of
isochrones published by the  Padua research
group\footnote{http://www. pleadi.astro.pd.it} and described in
the paper Girardi et al.(2000) The superposition of various
isochrones on the CMD shows the perfect coincidence of the
$8.913\times10^7$ yr isochrone for $Z=0.019$ with the cluster
diagram. The accuracy of age determining depends on the grid step
($\pm0.05logA$). Corresponding to this isochrone the turn-off
point of the MS is $(B-V)_0=-0.16$ mag.
\begin{figure} [t]
\centering{
\vbox{\psfig{figure=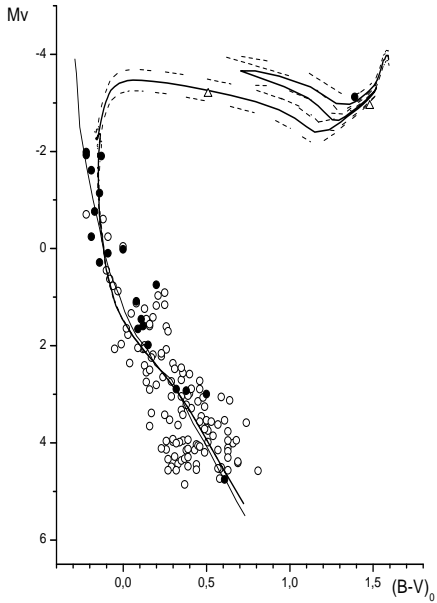,width=7cm,height=9cm}}\par}
\caption []{CMD diagram of all cluster members. ZAMS-
Schmidt-Kaler (1965). Isochrones: z=0.0019, solid line - $age=
8.913\times10^7$ yr; dot lines - adjacent ones. $\bullet$ - stars
supplied by photoelectric photometric data, $\circ$ - all the
other.}
\end{figure}
 The cluster luminosity function is plotted for 148 cluster members (Fig.\,7a).
As illustration the luminosity function for the field star is
presented here too. The  cluster mass function was plotted using
only stars of the cluster MS located lower than the turn-off point
$M_V=-0.41$ mag (Fig.\,7b). Star masses were calculated using the
data from the work of Girardi et al. (2000). The straight line
drawn  by the root-mean-square method with the 95\% confidence
level has the slope $\Gamma=-1.90\pm0.22$. That is in a good
agreement with the numerical results of Scalo (1998) in his
three-segment power law of the initial mass function for star
clusters and associations of Milky Way and LMC. The Trumpler\,2
mass function shows the lack of stars with the mass about
$2.8M_\odot$. This feature is observed also on the CMD for the
$M_V\sim+0.7$.
\begin{figure} [t]
\centering{
\vbox{\psfig{figure=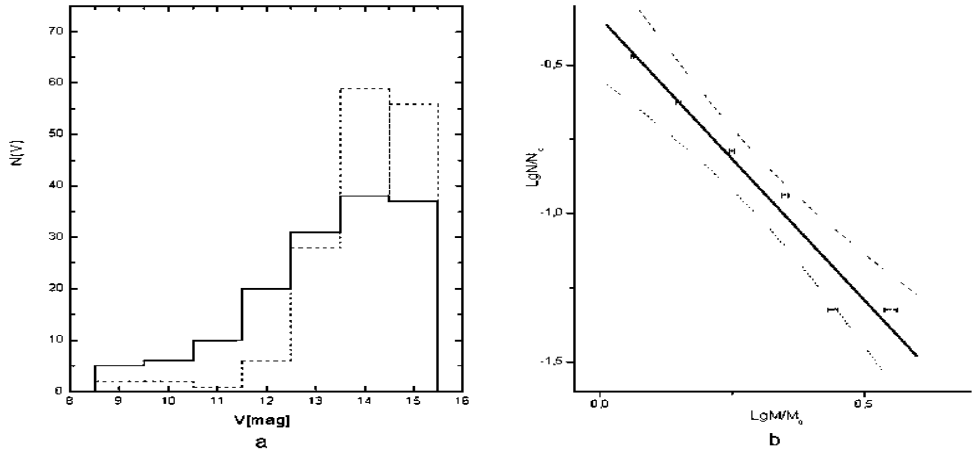,width=15cm,height=7cm}}\par}
\caption []{ a) Luminosity function and b) mass function of the
Trumpler\,2.}
\end{figure}

\section{Variable stars.}
In the studied  in the $80\arcmin\times80\arcmin$ area where star
proper motions were obtained eight variable stars from the the
General catalogue (GCVS) of Kholopov (1998) were found. All of
them are situated far from the center of the cluster
($R>15\arcmin$) and hardly belong to it in spite of the the fact
that they lie in 2$\sqrt{2}\sigma$ circle  on the VPD. Their data
are given in the part A of Table\,6.  Its rows contain 1 - the
GCVS star indication, 2 - the author's number, 3,4 - the right
ascention and the declination of the star, 5 - variability type, 6
- the distance from the center, 7,8 - $B,V$ magnitudes from
authors photometry catalogue(Table\,3), 9 - membership.
 \setcounter{table}{5}
\begin{table} [!]
\caption[]{Variable stars in the region of the open cluster
Trumpler\,2}
\begin{tabular}{lcccccccc}
\hline \multicolumn{9}{l}{\bf{A}}\\ Symbol  & LR Per & DX Per & DY
Per &  DZ Per &   VZ Per & EI Per & EK Per  &  EM Per\\ The
author's&&&&&&&&  \\ number & 2355  &   1856 & 1761 & 1571 & 996 &
1042 & 1381 & 513  \\ $\alpha (1950.0)$ &   02 29 14 & 02 31 14 &
02 31 42 & 02 32 22 & 02 35 07 & 02 36 15 & 02 36 35 & 02 37 52 \\
$\delta(1950.0)$ &  +56 16.1 & +56 05.0 & +55 55.6 & +55 55.9 &
+55 33.1 & +55 40.8 & +55 57 & +55 18.5\\ Variability type & SR: &
EA/SD: &  SRB &  EB/KE  & LB &  M & L & L
\\ R(arcmin) & 52.06  & 29.03 & 24.00 &  15.51 & 16.79  & 23.95  &
28.86 & 44.21
\\ Membership &prob & prob & prob & 93 & prob & prob& prob & prob \\
 B(mag)    &  15.39 &  15.49 &  12.69 &  14.11 &   14.20 &  14.35  &  15.34  &     15.51 \\
 V(mag)     &  14.16 &  14.92 &  12.38 &  12.10 &   8.94  &  13.21  &  14.70  &     10.40 \\

\end{tabular}
\begin{tabular}{lclllccccccc}
 \multicolumn{12}{l}{\bf{B}}\\

ID & The author's & MaScat & MedMag & MedEr & Ng & Quality &R &
Membership & B(mag) & V(mag) & S/N \\ & number &&&&&&&&& &\\
1803249 &  1137 & 0.155 & 12.38 & 0.056 & 49 & poor &   17.98 &
prob & 12.80 &   12.23 & 5
\\ 1803292 & 1139  &  0.147  &  12.145   &  0.044   &   44 & poor  &
17.37 &  prob  & 13.47  &   12.69   &  5   \\ 1803346 &  1143  &
0.2 & 12.654 &  0.068  &    42 & poor  &   18.13 &    prob &
13.27 & 12.65 & 5
\\ 1803553 &  1241  &  0.153  &  10.693   &  0.016   & 31 & poor  & 12.79
&  94  &  10.57  &   10.26   &  2   \\
 &&&&&&&&& 10.45  &   10.16 &  3 \\
 &&&&&&&&& 10.51 &  10.28 &  5 \\
1803662 & 1127 & 0.145 & 11.985  &  0.036 & 27& poor  & 11.68 & 93
& 12.11 & 11.63 & 2 \\
 &&&&&&&&& 11.81 & 11.75 & 3 \\
 &&&&&&&&& 11.95 & 11.71 & 5 \\
1803669 &  954 & 0.232 & 12.667 & 0.067 & 22& poor  & 15.05 & 94 &
13.30 & 12.86 &  5 \\ 1803717 & 1442 & 0.146 & 11.787 & 0.031 &
22& poor & 9.62 & 93 & 12.27 &  11.67 &  3 \\ &&&&&&&&& 12.26 &
11.47 & 4 \\ &&&&&&&&& 12.26 & 11.45 & 5 \\ 1803741 & 961 &  0.123
& 10.767 & 0.016 & 21 & poor & 10.13 & 94 &  10.47 & 10.26 & 1 \\
&&&&&&&&&  10.50 & 10.37 &  3 \\ &&&&&&&&&  10.42 & 10.27 &  4 \\
&&&&&&&&&  10.42 & 10.32 &  5 \\ 1921137  & 721 & 1.324 & 11.546 &
0.025 & 101 &good & 17.96 & prob & 12.79 & 12.19 & 3 \\ &&&&&&&&&
13.21 & 12.18 & 5\\ 1921548 & 1294 & 0.937 & 12.792 & 0.059 &
199&good & 10.55 & 93 & 13.35 & 12.66 & 4 \\ &&&&&&&&& 13.35 &
12.67 & 5 \\ 1922164 & 1393 & 0.62 & 13.622 & 0.115 & 205&high &
16.76 & prob & 14.41 & 13.82 & 5 \\ 1922260 & 1049 & 1.217 &
12.091 & 0.04 & 227 &high& 17.66 & prob & 14.41 & 12.12 & 5 \\
\hline
\end{tabular}
\end{table}
 The search of the variable stars was carried out in the recently
published database SKYDOT\footnote{http://skydot.lanl.gov} (SD)
(Wo\'zniak P.R. 2004)  relied on the Northen Sky Variability
Survey (NSVS). From the list of all the objects presented in the
SD we have selected only those ones which amplitude exceed for
more than 3 times the mean error. Twelve of them could be the
cluster members. The list of selected objects is given in the part
B of the Table\,6. Its columns contain 1 - the SD star indication,
2 - the author's number, 3 - the maximum brightness amplitude, 4 -
the mean magnitude, 5 - the mean error of the magnitude
determination, 6 - the number of the reliable observations during
the whole period, 7 - quality of observations, 8 - the distance
from the cluster center,9 - membership, 10,11 - $B,V$ magnitudes,
12 - the photomety specific number from the Table\,3. Star
magnitudes in the SD were determined in the natural system which
was close to V band. Curves of the brightness variation do not
permit to determine the types of the variability because of the
low quality of the most of the observations. Unfortunately the
technical reasons did not permit the investigators to observe the
bright stars and made impossible to compare their results with the
works of the other researchers. By the way this obstacle in
particular prevented us from the verification of our supposition
of the variability of the probable cluster member  star N 1045
(the red giant).

\begin{acknowledgements}
        We would like to thank  investigators of Pulkovo observatory
        N. Bronnikova, E. Polyakov, A.Devyatkin for their collaboration
in the observations. E. Jilinski thanks MCT Brazil for financial
support.
\end{acknowledgements}

{}

\end{document}